\documentclass{svproc}
\usepackage{url}

\usepackage{graphicx}        

\usepackage{amsmath}

\usepackage{wrapfig}

\begin{document}
\mainmatter              
\title{An analytical solution of the Social Force Model for uni-directional flow}
\titlerunning{An analytical solution of the SFM}  
%
\author{Tobias Kretz}
\authorrunning{T. Kretz} 
%
\tocauthor{Tobias Kretz}
\institute{PTV Group, D-76131 Karlsruhe, Germany\\
\email{Tobias.Kretz@ptvgroup.com}}

\maketitle              

\begin{abstract}
A function for the dependence of flow on pedestrian density is derived analytically from the 
Social Force Model (SFM) for the case of a homogeneous population walking in the same direction 
and being in steady state. Assuming that only nearest Voronoi neighbors effectuate forces the resulting 
function matches a variety of very different fundamental diagrams that were found empirically.
\keywords{pedestrian dynamics, fundamental diagram}
\end{abstract}
\section{Motivation}
\begin{figure}[ht!]
\centering
\includegraphics[width=\textwidth]{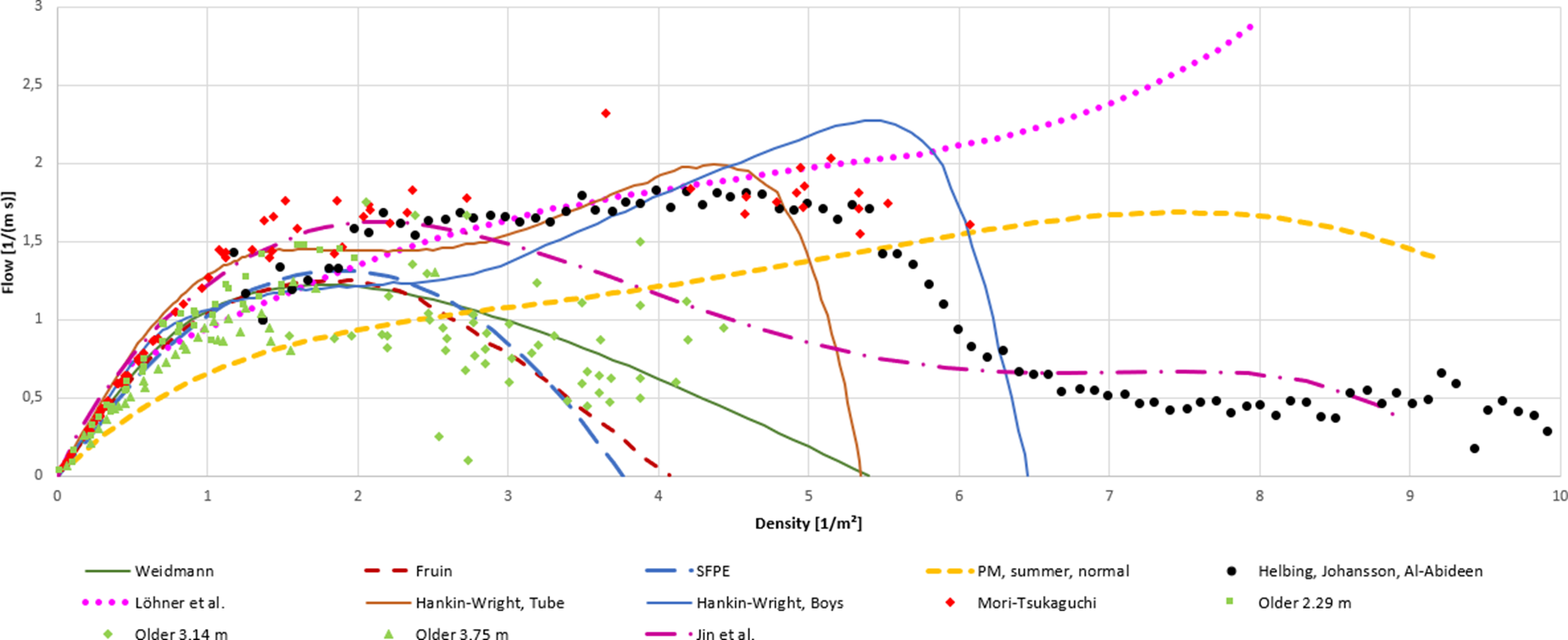} 
\caption{Empirical pedestrian fundamental diagrams as reported in \cite{weidmann1993transporttechnik,fruin1971pedestrian,dinenno2008sfpe,predtechenskii1978planning,helbing2007dynamics,lohner2018fundamental,hankin1958passenger,mori1987new,older1968movement,jin2017large}. Some of the data is publicly available at https://www.asim.uni-wuppertal.de/de/datenbank/data-from-literature/fundamental-diagrams.html. Where (mean value) functions are shown they are taken from literature, where no average function is given, the scatter plot is shown.}
\label{fig:PFD01}       
\end{figure}

The motivation for this study comes from different sources.
First, the variety of fundamental diagrams of pedestrian dynamics which were reported from field and laboratory -- as shown in Figure (\ref{fig:PFD01})  -- studies calls for an explanation. Is it measurement methods or evaluation techniques that causes these large differences? Is even the quality of a majority of these studies at question? Is the notion of a fundamental diagram for pedestrians -- dissenting with \cite{lighthill1955kinematic} -- simply not meaningful? Or is the variation a consequence of varying circumstances which then should be possible to be modeled with parameter value variation within one single model. This is emphasized when comparing with fundamental diagrams discussed for vehicular traffic flow -- for an overview see \cite{del1995functional} -- where there is also a bandwidth of shapes, but it is clearly not as wide as for pedestrian dynamics. Is this because the longer lasting research in vehicular dynamics has led to more agreement or can a reason be identified why pedestrian dynamics would yield different fundamental diagrams than vehicular dynamics?

Second, it was possible before to solve the SFM analytically for single file movement \cite{kretz2015social,kretz2016inflection2} which allowed to derive equations with which model parameter values can be computed from observables \cite{kretz2018some}. The latter is a decidedly valuable result since it allows direct calibration of the model (with all limitations following from approximations made as part of the analysis).

Third, an analytical solution of a simulation model allows to formulate an expectation for simulation results. This can be a helpful element for the verification of a software implementation.

\section{Setting, Assumptions, and Approximations}
For this purpose it is assumed that \dots
\begin{itemize}
\item[$\dots$] all pedestrians desire to walk in the same direction. This mainly defines the setting for which the obtained function for a fundamental diagram can (approximately) be valid. To a lesser degree it is also an approximation.
\item[$\dots$] the system is infinitely large or that there are periodic boundary conditions in both dimensions (walking on a torus / ''doughnut'') This is to avoid having to consider the effect from borders and walls and it implies that the solution will be a better approximation for large than for small systems.
\item[$\dots$] all pedestrians are identical in the sense that they are described with identical parameter values in the SFM, most importantly this means that the desired speed is set identical for all. This is essential for the analytical treatment since a discussion that includes parameter variations would be much more difficult. At the same time it is unrealistic and one has to be aware that even small variations may disturb and destroy a steady-state.
\item[$\dots$] only nearest Voronoi neighbors affect a pedestrian. Compared to the original SFM this is a strong limitation since there all pedestrians in a system produced a force on any other. The latter may be largely unrealistic, but it is plausible that in reality beyond nearest also next to nearest and even further neighbors can have a direct effect.
\item[$\dots$] the system is in a steady-state, all forces on a pedestrian cancel to zero. It is assumed that a steady-state exists and is stable.
\item[$\dots$] pedestrians walk in a certain formation, namely on the grid points of a triangular lattice, which is not necessarily regular, but can be stretched or quenched. It is not clear to which degree this is an approximation. With the exception of \cite{porzycki2014pedestrian} there is little research on walking formation or just the number of Voronoi neighbors.
\item[$\dots$] density changes affect only longitudinal but not lateral spacing. This is an approximation, but the tendency appears to exist. A bounding box which includes nearest or next to nearest neighbors in data from a laboratory experiment \cite{JSC2018pedestrian} typically has a higher width to length ratio for higher than for lower densities and this makes sense since with speed stride length grows which extends space requirements in walking direction, but not transverse.
\item[$\dots$] the analytically derived function for steady-state matches (approximately) the average function of empirical data. This is not relevant for the analytical treatment as such, but for the comparison of analytical function and empirical average it is of course important to bear in mind that strictly speaking two different properties are being compared.
\end{itemize} 

\section{Analysis and Results}
\begin{figure}[ht!]
\centering
\includegraphics[width=0.3\textwidth]{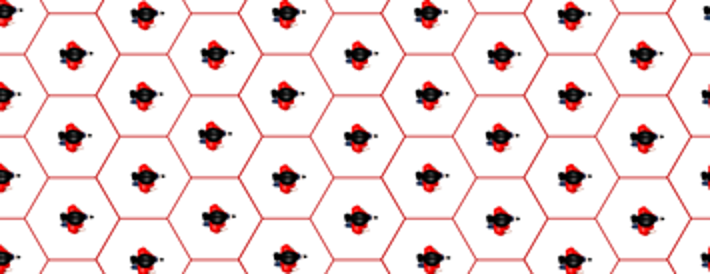} \hspace{10pt}
\includegraphics[width=0.3\textwidth]{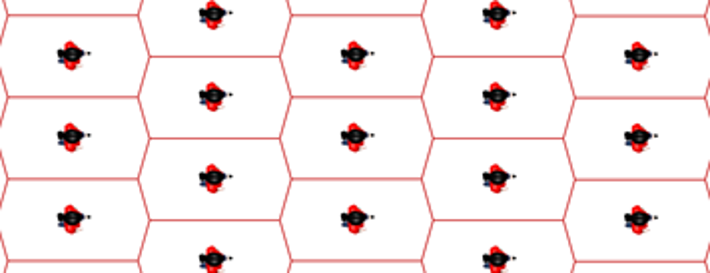} \hspace{10pt}
\includegraphics[width=0.3\textwidth]{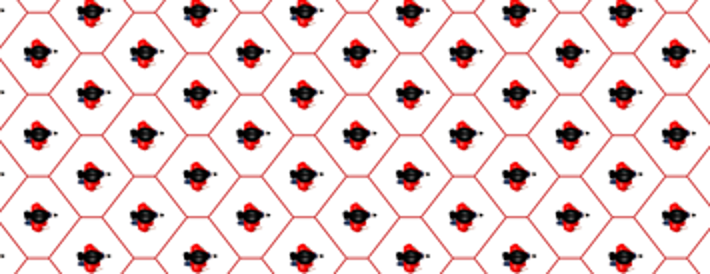} 
\caption{Hypothetical walking formation at medium density, when Voronoi cells are regular hexagons (left), lower density (center) and higher density (right). The lateral spacing is the same in all three cases (i.e. the number of pedestrians in a row remains the same). Walking direction is $\rightarrow$ .} 
\label{fig:densities}      
\end{figure}

Together with these assumptions a highly symmetric walking formation is assumed as shown in Figure (\ref{fig:densities}).
The task is now to compute a speed for each density based on the geometry of the walking formation and some appropriate model of pedestrian dynamics, here the SFM serves for this purpose, to be precise circular specification or elliptical specification II (either will work) as of \cite{johansson2007specification}.

\begin{figure}[ht!]
\centering
 $\vcenter{\hbox{\includegraphics[width=0.16\textwidth]{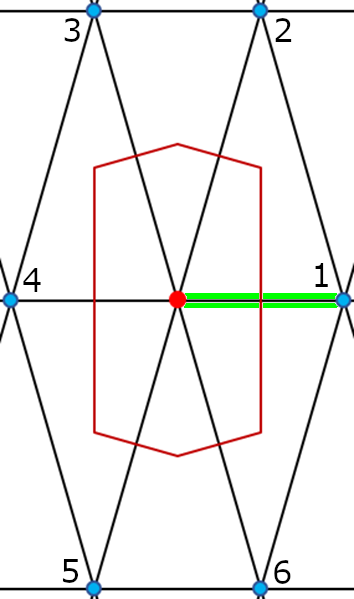}}} \hspace{6pt}
 \vcenter{\hbox{\includegraphics[width=0.16\textwidth]{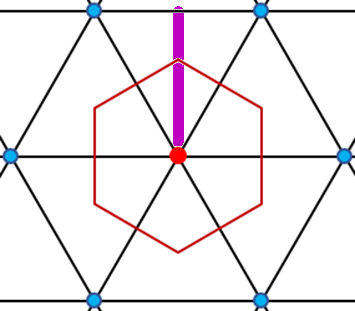}}} \hspace{6pt}
 \vcenter{\hbox{\includegraphics[width=0.16\textwidth]{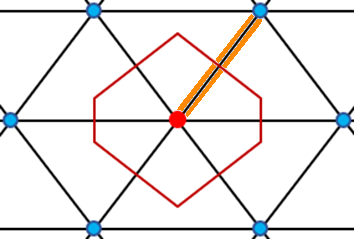}}} \hspace{6pt}
 \vcenter{\hbox{\includegraphics[width=0.16\textwidth]{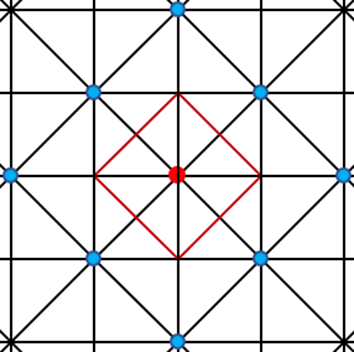}}} \hspace{6pt}
 \vcenter{\hbox{\includegraphics[width=0.08\textwidth]{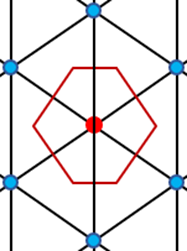}}} \hspace{6pt}
 \vcenter{\hbox{\includegraphics[width=0.08\textwidth]{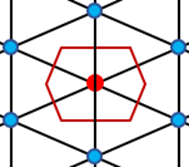}}}$\\
\caption{Voronoi cells from lower (left) to higher (right) densities. Walking direction is $\uparrow$. Comparing the first and the third figure one can easily see that the Voronoi cell is not simply stretched and quenched with the grid of the walking formation. This has an important consequence: at lower densities a pedestrian has only lateral and diagonal neighbors, whereas above a density of $\rho=2/b^2$ (4th subfigure) a pedestrian has diagonal and longitudinal neighbors. In the left subfigure neighbors are numbered for use in equation (\ref{eq:simpleSFM}) and the green line marks the lateral spacing $b$. In the second subfigure the height $h$ of the Voronoi cell is highlighted in magenta and in the third subfigure the 'diagonal' $d$ is wrapped orange. }
\label{fig:Voronoi}      
\end{figure}

As a first step it is interesting to note that regarding Voronoi neighbors there are two regimes: one for low and one for high densities as Figure (\ref{fig:Voronoi}) visualizes. 

This is a consequence from the Voronoi cells not simply stretching and quenching along with the walking formation, but its edges transform conformal with connecting lines between nearest neighbors as can be seen in Figure (\ref{fig:Voronoi}). This alternative transformation behavior of the cells is neutral with respect to area content and thus has no effect on how density is computed, namely as  
\begin{equation}
\rho=\frac{1}{bh}
\end{equation}
with $b$ and $h$ defined as depicted in Figure (\ref{fig:Voronoi}).

It is clear that forces from lateral neighbors (numbers 1 and 4 in Figure (\ref{fig:Voronoi})) cancel and so do -- as a consequence of the high symmetry -- pairwise all contributions to the lateral force component. This makes the lateral spacing $b$ an input parameter to the model. The actual walking direction matches the desired walking direction and there is no component of velocity or acceleration orthogonal to it. Thus, we are faced essentially with a 1d equation of motion.

The full model (circular specification)

\begin{eqnarray}
\ddot{\vec{x}}_i(t) = \frac{\vec{v}_{0,i}-\dot{\vec{x}}_i(t)}{\tau_i} + \tilde{A}_i \sum_j w(\vec{x}_i(t),\vec{x}_j(t),\dot{\vec{x}}_i(t),\lambda_i)e^{-\frac{|\vec{x}_j(t)-\vec{x}_i(t)|-R_i-R_j}{B_i}}\hat{e}_{ij}\\
w(\vec{x}_i(t),\vec{x}_j(t),\dot{\vec{x}}_i(t),\lambda_i) = \lambda_i + (1-\lambda_i) \frac{1+\cos(\theta_{ij}(\vec{x}_i(t),\vec{x}_j(t),\dot{\vec{x}}_i(t)))}{2}
\end{eqnarray}

can be simplified and re-arranged as a consequence of the high symmetry and the steady-state if only nearest Voronoi neighbors contribute to the force:

\begin{eqnarray}
\dot{x}&=&\vec{v}_{0}+\tau A \left(\sum_{j=5,6} \cos(\theta_j) w(\lambda,\theta_j) e^{\frac{-d}{B}} - \sum_{j=2,3} \cos(\theta_j)w(\lambda,\theta_j) e^{\frac{-d}{B}}\right) \label{eq:simpleSFM}\\
\dot{x}&=&\vec{v}_{0} - 2\tau A (1-\lambda) \frac{h^2}{d^2} e^{\frac{-d}{B}}
\end{eqnarray}

and get as speed-density relation and fundamental diagram

\begin{equation}
j=v_{0}\rho \left(1 - 2 (1-\lambda) A\frac{\tau}{v_0} \frac{1}{1+\frac{\rho^2b^4}{4}} e^{-\frac{1}{2}\frac{b}{B}\sqrt{1+\frac{4}{b^4\rho^2}}}\right) \label{eq:fd}
\end{equation}

This holds for the geometry as depicted in the first three subfigures of Figure (\ref{fig:Voronoi}) which is the low density range. For the high density range the analysis a term from the longitudinal neighbor (the one between and ahead of neighbors number 2 and 3 in Figure (\ref{fig:Voronoi})) must be added which can be calculated accordingly.

\begin{figure}[ht!]
\centering
\includegraphics[width=0.3\textwidth]{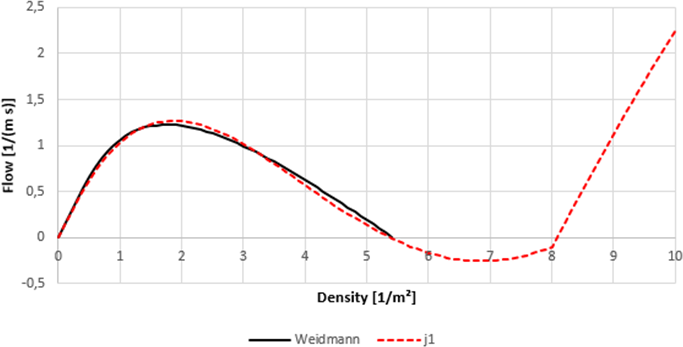} \hspace{10pt}
\includegraphics[width=0.3\textwidth]{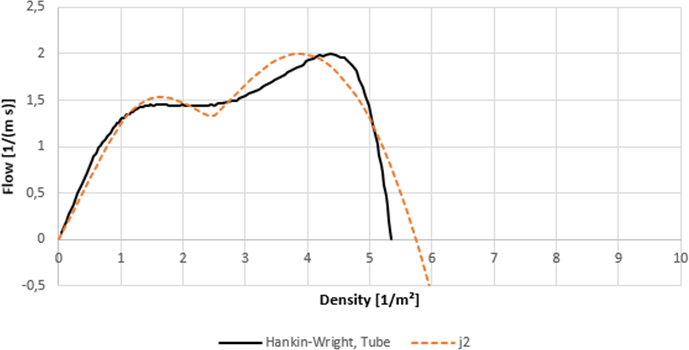} \hspace{10pt}
\includegraphics[width=0.3\textwidth]{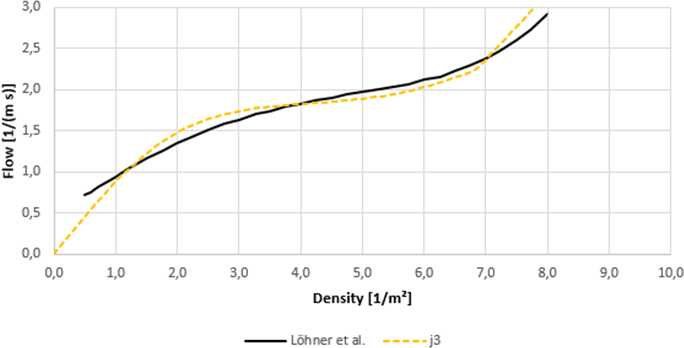} \\ \vspace{10pt}
\includegraphics[width=0.3\textwidth]{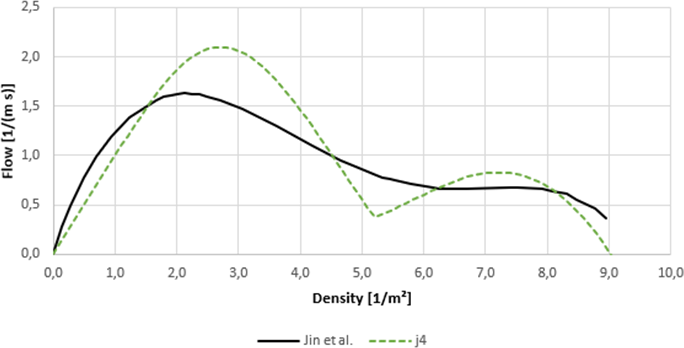} \hspace{10pt}
\includegraphics[width=0.3\textwidth]{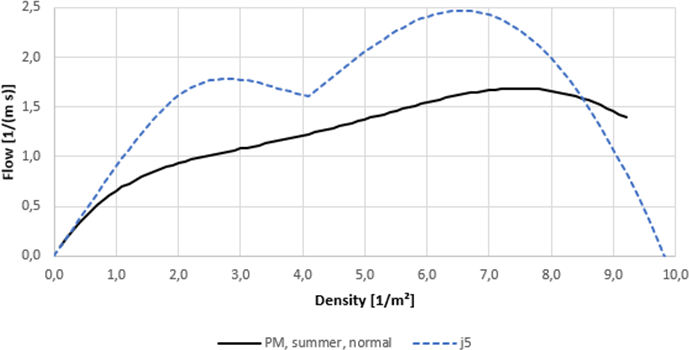}
\caption{Empirical fundamental diagrams (Weidmann, Hankin-Wright, L{\"o}hner et al., Jin et al, Predt.-Mil.) compared to equation (\ref{eq:result}) with parameters from Table \ref{tab:parameters}.} 
\label{fig:FD}      
\end{figure}

Finally, the two ranges have to be connected such that a steady function follows. This comes down to the question if the longitudinal neighbor should have a full effect if it shares only one point or a very short edge with the pedestrian for whom forces are computed. A straightforward idea is to weigh forces with the length of the shared edge relative to the total circumference of the Voronoi cell. With this, the function for the fundamental diagram over the full density range results as 

\begin{eqnarray}
j(\rho) &=& \rho v_0 \left(1 - \alpha \left(2 g_{diag} \frac{1}{1+k^2} e^{-\frac{1}{2}\frac{b}{B} \sqrt{1+\frac{1}{k^2}}} + g_{long} e^{-\frac{b}{B}\frac{1}{k}}\right)\right) \label{eq:result}\\
g_{diag} &=& \frac{4l_{diag}}{4 l_{diag} + 2 l_{long} + 2 l_{lat}} \text{;   } g_{long} = \frac{2l_{long}}{4 l_{diag} + 2 l_{long} + 2 l_{lat}}\\
l_{diag} &=& \frac{b}{2} \sqrt{1+k^2}\min\left(1, \frac{1}{k^2}\right) \text{;   } l_{long} = \frac{b}{2} \max\left(0, 1-\frac{1}{k^2}\right)\\
l_{lat}  &=& \frac{b}{2} \max\left(0, \frac{1}{k} - k\right)\\
\alpha &=& (1-\lambda) A \frac{\tau}{v_0}\text{;   } k := \rho \frac{b^2}{2}
\end{eqnarray}
where an $l$ gives the length of the respective edge and a $g$ a weight factor. 
\begin{wraptable}[9]{r}{0.5\linewidth}
\centering
\vspace{-2\baselineskip}
\begin{tabular}{l|rrrrr}
\hline
              & P1 & P2 & P3 & P4 & P5 \\
\hline
$v_0$ [m/s]   & 1.3 & 1.3 & 0.9 & 1.0 & 0.9 \\
$A$ [m/s$^2$] & 4.2 & 34 & 3 & 35 & 22 \\
$B$ [m]       & 2.5 & 0.23 & 0.6 & 0.168 & 0.18 \\
$b$ [m]       & 0.5 & 0.9 & 0.54 & 0.62 & 0.7 \\
\hline
\end{tabular}
\caption{Parameter values which allow to approximate various empirical fundamental diagrams. It is $\lambda=0.1$ and $\tau=0.4$ s for all.}
\label{tab:parameters}
\end{wraptable}
$\alpha$ is a characteristic value for a parametrization of the SFM since it relates the forward-backward asymmetry $(1-\lambda)$, the base strength of forces between pedestrians $A$ and the base self-propelling acceleration $v_0/\tau$. $k$ is a scaled density which has the value $k=1$ at the border of low and high density range. With the parameters from Table \ref{tab:parameters} equation (\ref{eq:result}) allows to reproduce fundamental diagrams of all shapes as shown in Figure \ref{fig:FD}.

\section{Summary -- Conclusions -- Outlook}
With some simplifying assumptions it was possible to derive analytically from the SFM a function for the fundamental diagram which allows to approximate the entire variety of documented empirical fundamental diagrams by means of parameter variation. This does not guarantee, but it is an indication, that all of the empirical data is valid, despite it being seemingly contradicting, and that the SFM with Voronoi neighborhoods is a realistic model of pedestrian dynamics. 

One of the obvious next steps is to implement a variant of the SFM with a Voronoi neighborhood as source of inter-pedestrian forces. 

While the starting point was the Social Force Model along the course of the analytical treatment its details seemed less and less relevant. As long as the interaction between pedestrians decreases monotonically with distance and as long as there is some suppression of effects from behind, the main results will not change significantly. One could almost say ''the neighborhood is the model''. Consequently, an empirical investigation into the structure and geometry of Voronoi cells in real systems appears to be very interesting, particularly into how realistic the assumption of a constant lateral spacing is. 
%
%
\bibliographystyle{spphys} 
\bibliography{TGF19a} 
\appendix
\section{Supplemental Material and Information related to Section 'Motivation'}
\begin{figure}[ht!]
\centering
\includegraphics[width=0.3\textwidth]{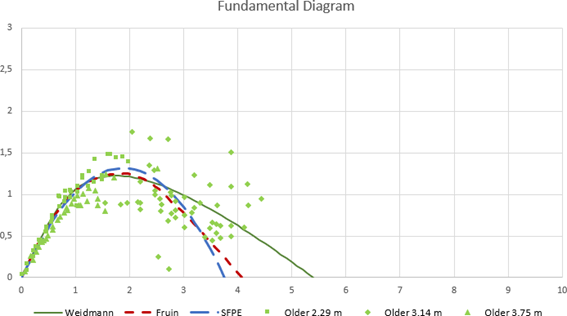} \hspace{10pt}
\includegraphics[width=0.3\textwidth]{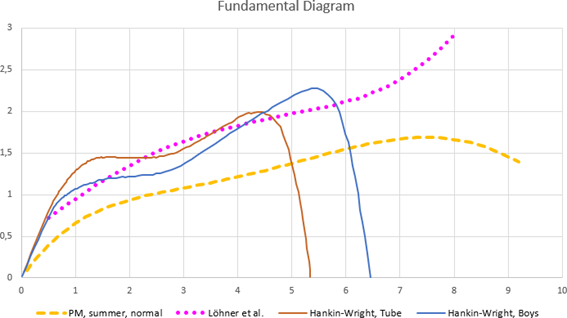} \hspace{10pt}
\includegraphics[width=0.3\textwidth]{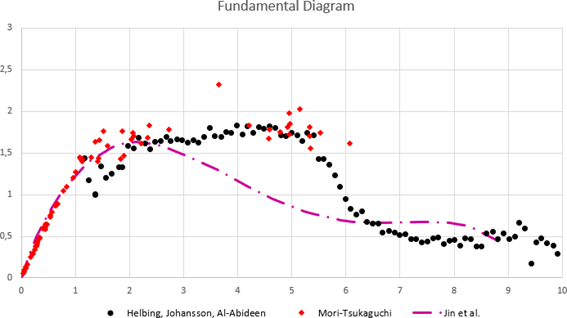} 
\caption{Three categories of fundamental diagrams, referred to here as 'left leaning, simple', 'right leaning', and 'left leaning, complex'.} 
\label{fig:FDcat}      
\end{figure}

The fundamental diagrams shown in Figure \ref{fig:FD} can be categorized in three groups as shown in Figure \ref{fig:FDcat}.

These three groups can be characterized as follows:

'left leaning, simple' (includes Weidmann, Fruin (and with it SFPE) as well as Older):
\begin{itemize}
\item a moderate maximum density $\rho_{max} \approx< 6/m^2$, 
\item a moderate capacity $j_c \approx < 1.5/m/s$,
\item left leaning, i.e. the density at capacity is smaller than half of the maximum density $\rho_c < \rho_{max}/2$,
\item a comparably simple functional shape, no inflection point, negative curvature $d^2j/d\rho^2 < 0$ on the entire range of densities.
\end{itemize}

'right leaning' (includes Predtechenskii and Milinskii, L{\"o}hner et al., as well as Hankin and Wright)
\begin{itemize}
\item a maximum density $\rho_{max}$ above $5/m^2$, possibly beyond $10/m^2$, 
\item a comparably high capacity $j_c$ above $1.5/m/s$ reaching to values close to $3/m/s$,
\item right leaning, i.e. the density at capacity is larger than half of the maximum density $\rho_c > \rho_{max}/2$,
\item two inflection points at lower densities than capacity.
\end{itemize}

'left leaning, complex' (includes Helbing et al., Mori and Tsukaguchi as well as Jin et al.):
\begin{itemize}
\item a high maximum density $\rho_{max} \approx 10/m^2$, 
\item a moderately high capacity $j_c \approx 1.7/m/s$,
\item left leaning, i.e. the density at capacity is smaller than half of the maximum density $\rho_c < \rho_{max}/2$,
\item aside the maximum which marks capacity it appears that there is a second, weakly pronounced local maximum at high densities ($\approx 8.5/m^2$) and between the two maximums an equally weak pronounced minimum.
\end{itemize}

For this grouping it is first assumed that in the situation where L{\"o}hner et al. measured even higher densities could exist, but would soon lead to a collapse of the flow. Second, it is assumed that in the situation where Mori and Tsukaguchi measured for even higher densities flow would evolve similarly as in the case of Helbing et al..

The functions as presented here, have their immediate source here
\begin{itemize}
\item Weidmann analytical function from {\em Transporttechnik der Fu{\ss}g{\"a}nger}, i.e. the Kladek\footnote{H. Kladek {\em {\"U}ber die Geschwindigkeitscharakteristik auf Stadtstra{\ss}enabschnitten} PhD Thesis (1966)} formula which has been applied before by Newell\footnote{G.F. Newell {\em Nonlinear effects in the dynamics of car following} Operations Research (1961)}. The empirical function given by Weidmann has a slightly different shape.
\item Fruin graphically grasped from chart in {\em Pedestrian Planning and Design}. 
\item Hankin and Wright, Helbing, Johansson, and Al-Abideen, Mori and Tsukaguchi are the data made available on https://www.asim.uni-wuppertal.de/ (given in speed vs. density).
\item Predtechenskii and Milinskii numerically copied from the table in the appendix of the German edition Personenstr{\"o}me in Geb{\"a}uden of their book. One or two apparently transposed digits in the source were corrected.
\item L{\"o}hner et al. graphically grasped from Figure 13 in {\em Fundamental diagrams for specific very high density crowds}. 		
\item Jin et al graphically grasped from Figure 3 in {\em Large-scale pedestrian flow experiments under high-density conditions}.
\item where sources are given as speed(density) flow has been computed as product of speed and flow.
\end{itemize}

\begin{figure}[ht!]
\centering
\includegraphics[width=\textwidth]{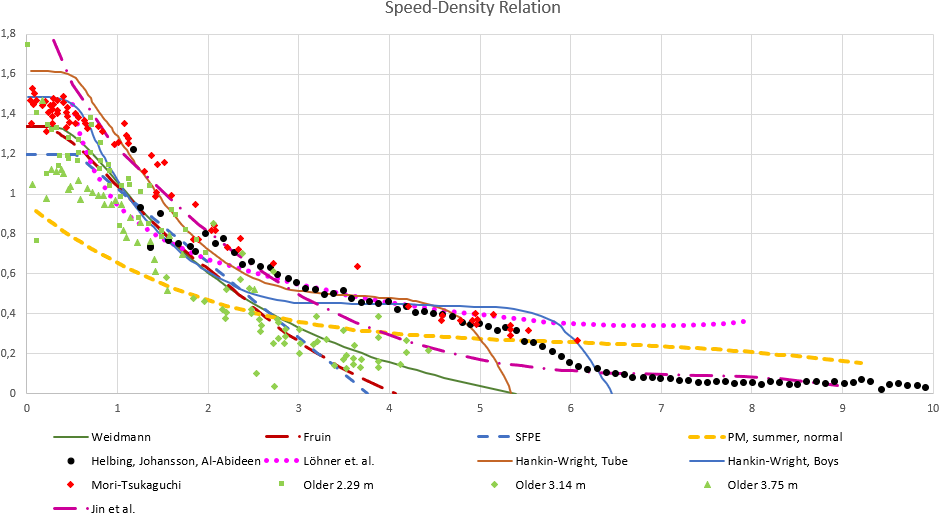} 
\caption{Speed-density relations belonging to fundamental diagram of Figure \ref{fig:FD}.} 
\label{fig:speed-density}      
\end{figure}

The speed-density diagrams belonging to the fundamental diagrams of Figure \ref{fig:FD} are shown in Figure \ref{fig:speed-density}. For these no categorization is apparent. Only the speed-density relation from L{\"o}hner et al. stands out with its minimum.

\begin{figure}[ht!]
\centering
\includegraphics[width=\textwidth]{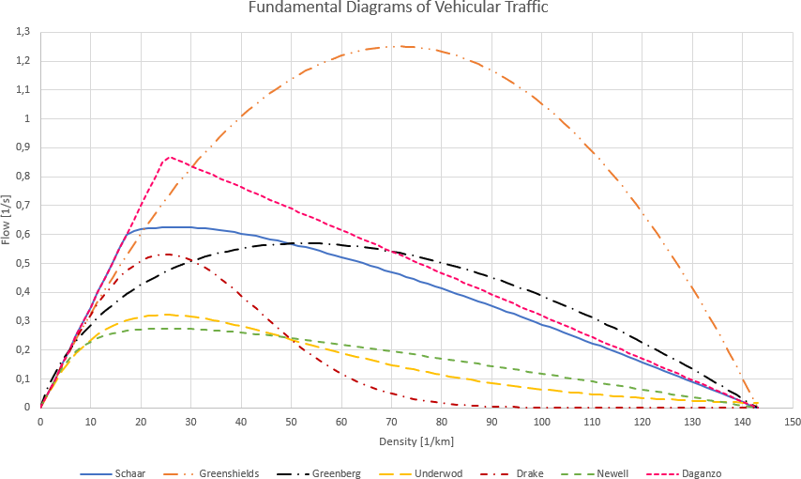} 
\caption{Fundamental diagrams of vehicular traffic. The original functions have been re-parametrized to have a maximum density of 143/km and (where degrees of freedom allowed) capacity at a density of 25/km and (where degrees of freedom still allowed) capacity at 0.5/s.} 
\label{fig:FDveh}      
\end{figure}

For comparison Figure \ref{fig:FDveh} shows a selection of well-known fundamental diagrams of vehicular traffic. It can be seen that the functional shape varies less than for pedestrian dynamics.

Concerning the second stated motivation aspect it might be interesting to add that the work flow that led to the result presented in this contribution was done before for single-file movement in the referenced works. The resulting analytical functions for fundamental diagrams and comparison of results for 'nearest neighbors only' and 'all pedestrians to infinity' interaction with empirical data gave a clear indication that the former is more realistic. The resulting fundamental diagram is
\begin{eqnarray}
j &=& \rho v_0 \left( 1 - \alpha \frac{1}{1+\kappa e^{-\frac{1}{B\rho}}}e^{-\frac{1}{B\rho}} \right) \\
\alpha &=& (1-\lambda) A \frac{\tau}{v_0}
\end{eqnarray}
with $0\leq\kappa\leq1$ suppressing forces according to neighborhood degree (nearest neighbor $\kappa^0$, next to nearest neighbor $\kappa^1$, and so on), which means that with $\kappa=0$ only nearest neighbors interact and $\kappa=1$ is the original Social Force Model. For $\kappa=0$ this is the fundamental diagram proposed independently by Newell and later Kladek and used by Weidmann to approximate the fundamental diagram of pedestrians if $B\rho_{max}\ln(\alpha) = 1$. Also for $\kappa=0$ the equation for the fundamental diagram can be solved for the model parameters $\alpha$ and $B$ depending explicitly on the observables capacity $j_c$, stand-still density $\rho_{max}$ and free speed $v_0$:
\begin{eqnarray}
\alpha &=& \left[ -W\left( -\frac{1-q}{e}\right)\frac{e}{1-q}\right]^{\frac{q}{1-q}}\\
B &=& -\frac{1}{W\left( -\frac{1-q}{e}\right)}\frac{1}{\rho_{max}}\frac{1-q}{q}\\
q &=& \frac{j_c}{v_0\rho_{max}}
\end{eqnarray}
with $W()$ being the Lambert W-function\footnote{WolframMathWorld {\em Lambert W-Function}}.

Such an explicit function of model parameters depending on observable, empirical values is the easiest situation for calibration. Once the empirical values are obtained the model parameter values can simply be calculated such that the model reproduces the observed data (approximately, since to obtain the analytical results a number of simplifying assumptions had to be made; still one can hope that a good starting set of parameters can be computed). This was a great motivation to look for an analytical solution beyond single-file movement. While such a solution could be obtained, regrettably, the function does not allow to solve it analytical for model parameters.

\section{Supplemental Material and Information related to Section 'Settings, Assumptions, and Approximations'}
\begin{figure}[ht!]
\centering
\includegraphics[width=0.612\textwidth]{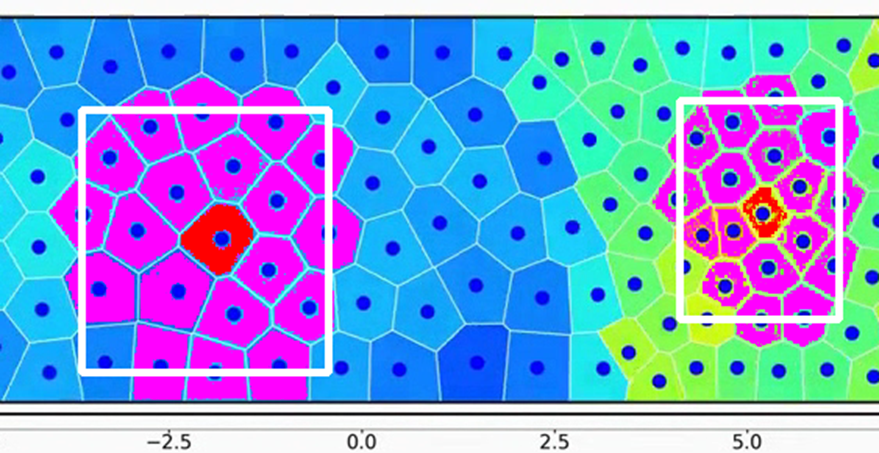} 
\caption{Voronoi tessellated snapshot from a laboratory experiment (edited from the last frame of video youtube.be/WBBJU2meS34). There is a high density to the right and a low density to the left. Walking direction is from left to right. The edges of the rectangles are oriented in walking direction and orthogonal to it. The rectangles are the smallest rectangles which include the coordinates of all nearest and next to nearest neighbors (which all are marked magenta). } 
\label{fig:Vf}      
\end{figure}

Concerning the fixed lateral spacing an example which exhibits at least the tendency is the last frame in a video from J{\"u}lich Supercomputing Center uploaded at youtube\footnote{{\em Voronoi Density}, https://www.youtube.com/watch?v=WBBJU2meS34, channel {\em Pedestrian and Fire Dynamics} maintained by the division Civil Security and Traffic at J{\"u}lich Supercomputing Centre, contact person: M. Chraibi}): Figure \ref{fig:Vf} shows two regions with clearly different density: higher to the right, lower to the left. In each zone one pedestrian has been chosen as 'center' of the smallest rectangle which encloses all nearest and next to nearest neighbors.

The low density rectangle covers an area of 10.86 m$^2$ with a length of 3.19 m and a width of 3.40 m (ratio 1:1.06). The high density rectangle covers 5.89 m$^2$ with a length of 2.07 and a width of 2.84 m (ratio 1:1.38). If the lateral spacing were exactly constant -- with the low density rectangle as a reference -- the high density rectangle would have to have a length of 1.73 m and a width of 3.40 m (ratio 1:1.97). Thus, in this example longitudinal spacing changes more with density than lateral spacing. Setting lateral spacing constant appears to be a justified albeit coarse approximation.

\begin{figure}[ht!]
\centering
\includegraphics[width=0.5\textwidth]{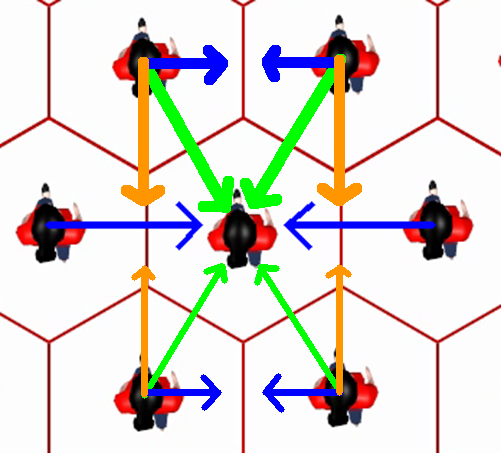} 
\caption{Sketch of forces. Absolute force value is displayed with thickness of arrows. For each lateral (blue) force there is an equally strong force in opposite direction. For longitudinal forces (orange) there is a none zero net force which balances with the driving force. } 
\label{fig:forces}      
\end{figure}

Saying the lateral spacing is fixed does not say to which value. Since inter-pedestrian forces in lateral direction all cancel out due to the high symmetry -- see Figure \ref{fig:forces} -- the value for lateral spacing cannot result from the model, but takes the role as input parameter into the model. In a simulation scenario which models a real situation (no periodic boundaries or infinite model size) this is different. The lateral spacing - be it walking next to each other in a near steady-state or during an overtaking process -- between one pedestrian and another is computed by the model.

\section{Supplemental Material and Information related to Section 'Analysis and Results'}
\begin{figure}[ht!]
\centering
\includegraphics[width=0.45\textwidth]{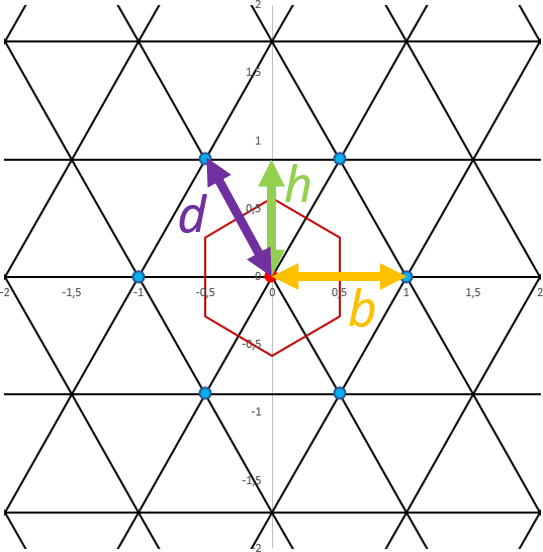} \hspace{11pt}
\includegraphics[width=0.45\textwidth]{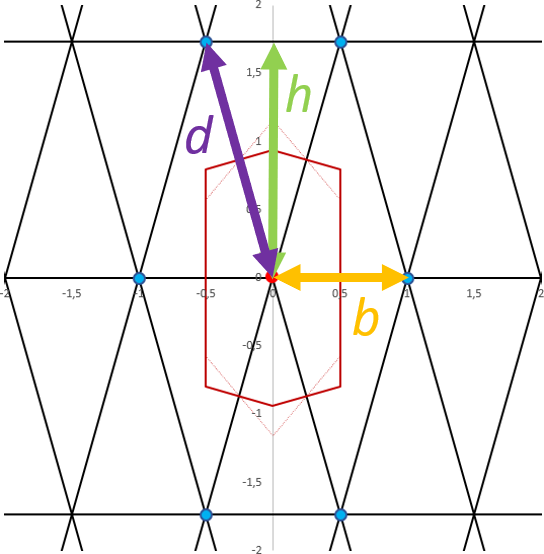}
\caption{Walking formation for a higher (left) and a lower (right) density. The Voronoi cells are marked in red. If the Voronoi cell from the left side were scaled with the walking formation it would have the shape as the slim red line in the right side figure indicates. } 
\label{fig:nomenclaturaB}      
\end{figure}

Figure \ref{fig:nomenclaturaB} shows the walking formation with the Voronoi cells and geometric definitions of quantities $b$, $h$, and $d$. The angle $\theta$ that appears in the $w_\lambda(\cos(\theta))$ function is between the lines marked for $h$ and $d$.

The Social Force Model needs as input the values of $d$ and $\cos(\theta)$. Trivially,
\begin{eqnarray}
d^2 &=& h^2 + \frac{b^2}{4}\\
\cos(\theta) &=& \frac{h}{d}
\end{eqnarray}

The area content of the Voronoi cell is
\begin{equation}
A = b h
\end{equation}
which means that walking density is
\begin{equation}
\rho=\frac{1}{bh}.
\end{equation}
That means that $h$ can be expressed through $\rho$ and $b$, and with it as well $d$ and $\cos(\theta)$:
\begin{eqnarray}
d^2 &=& \frac{1}{(b\rho)^2} + \frac{b^2}{4}\\
\cos(\theta) &=& \sqrt{\frac{1}{1+\frac{\rho^2b^4}{4}}}
\end{eqnarray}

\begin{figure}[ht!]
\centering
\includegraphics[width=0.612\textwidth]{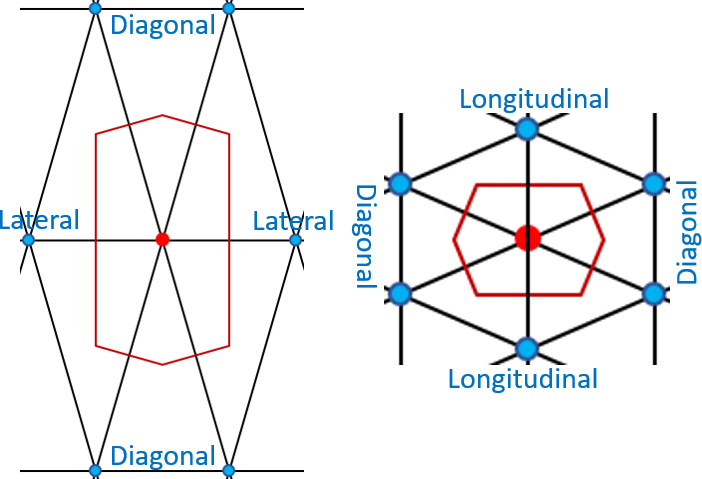}
\caption{Naming of neighbors for low (left) and high (right) densities. } 
\label{fig:nomenclaturaC}      
\end{figure}

As noted above, all inter-pedestrian force components orthogonal to the direction of motion (the direction of desired velocity) cancel out. The force component in direction of motion from a particular pedestrian scales with another $\cos(\theta)$ with the full force from that pedestrian. The four diagonal pedestrians (see Figure \ref{fig:nomenclaturaC} for naming convention of neighbors) together produce a scaling with a factor $2(1-\lambda)\cos(\theta)$. Together these factors result in the fundamental diagram of equation (\ref{eq:fd}).

For large densities $\rho>2/b^2$ there is an additional Voronoi neighbor walking straight ahead (''longitudinal neighbor'') and one straight behind. In this case it is 
\begin{eqnarray}
d &=& 2 h = \frac{2}{b\rho}\\
\cos(\theta) &=& 1
\end{eqnarray}
The lateral neighbors do not exist in this case, yet this has no effect since their effect canceled out anyway for small densities.

In the transition from low to high density the longitudinal neighbor is ''suddenly'' a Voronoi neighbor. This raises the question if the force from longitudinal neighbors should act fully from this point or if it should be damped in one way or another. Without smoothing the higher density range has an additional force term which means that speed and flow will drop at $\rho=2/b^2$. Considering that for vehicular traffic a number of discontinuous fundamental diagrams have been proposed, this would not necessarily have to be unrealistic. However, the empirical data does not suggest a discontinuous flow-density or speed-density function. Therefore it makes sense to investigate into a number of simple weighing functions, simple, because the data does not allow to substantially distinguish between subtle differences between more advanced functions. These functions or rules have been considered:

\begin{itemize}
\item {\bf Only diag}: This weighing rule ignores longitudinal neighbors for the whole range of densities, even if they belong to the set of Voronoi neighbors at high densities. This would violate the basic idea of this contribution, thus it is only included for comparison.
\item {\bf Always long}: In this rule the contribution of longitudinal neighbors is always considered even at low densities where they do not belong to the set of Voronoi neighbors.
\item {\bf Step}: This is the 'no weighing' rule. At low densities there is no longitudinal neighbor, at high densities they are considered without additional factor.
\item {\bf Long rel diag}: The contribution of longitudinal neighbors is scaled down with a factor that matches the ratio of the length of common Voronoi edges between the central pedestrian and a longitudinal neighbor and between the central pedestrian and a diagonal neighbor (but only up to a factor of 1).
\item {\bf Share}: Each neighbor contributes with a factor according to the share of its common edge with the central neighbor in the Voronoi cells circumference, but without the contribution of lateral neighbors.
\item {\bf Circum}: Each neighbor contributes with a factor according to the share of its common edge with the central neighbor in the Voronoi cells circumference.
\end{itemize}

\begin{figure}[ht!]
\centering
\includegraphics[width=0.612\textwidth]{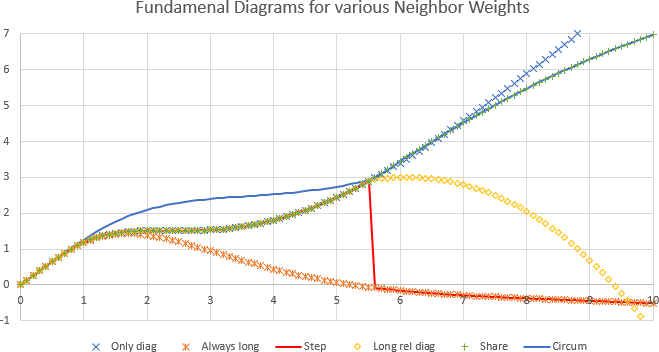}
\caption{Fundamental diagrams which differ only by neighbor weights. Parameter values are: $v_0=1.3$ m/s, $\tau=0.4$ s, $A=5$ m/s$^2$, $\lambda=0.1$, $B=0.5$ m, $b=0.6$ m.} 
\label{fig:weights}      
\end{figure}

The effect of these weighing structures on the fundamental diagram is shown in Figure \ref{fig:weights}. 

At this point it is important to recall that part of the motivation was to make use of these results for micro simulations. In micro simulations there is no exact steady-state and the walking formation never has exactly the proposed form. It is therefore never possible to identify diagonal, lateral, and longitudinal neighbors definitely. The weighing rules, however, depend on an identification of neighbor category, except for ''Step'' and ''Circum'' and of these only ''Circum'' has a continuous fundamental diagram. Therefore ''Circum'' is chosen as weighing rule and function.

\begin{figure}[ht!]
\centering
$\vcenter{\hbox{\includegraphics[width=0.3\textwidth]{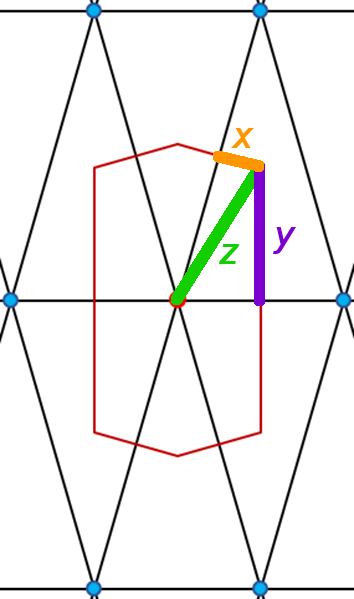}}} \hspace{11pt}
\vcenter{\hbox{\includegraphics[width=0.3\textwidth]{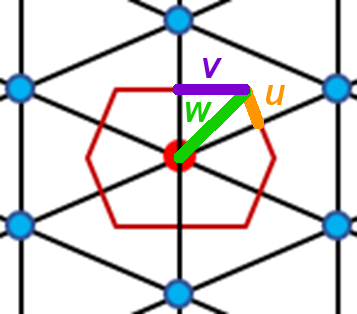}}}$
\caption{Calculating edge lengths with the help of $x=l_{diag}/2$, $y=l_{lat}/2$, and $z$, as well as $u=l_{diag}/2$, $v=l_{long}/2$, and $w$.}
\label{fig:edges}      
\end{figure}

\begin{figure}[ht!]
\centering
$\vcenter{\hbox{\includegraphics[width=0.45\textwidth]{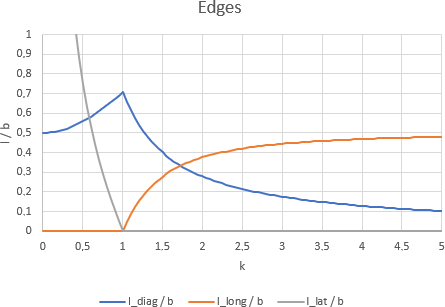}}} \hspace{11pt}
\vcenter{\hbox{\includegraphics[width=0.45\textwidth]{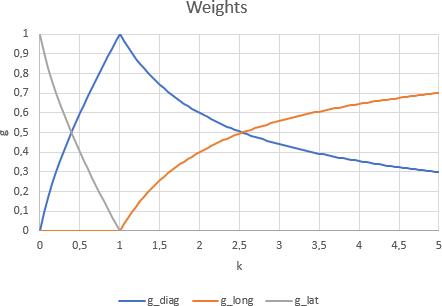}}}$
\caption{Edge lengths as multiples of $b$ in dependence of $k$ (left). Resulting weights for ''Circum'' weighing (right).}
\label{fig:edgesweights}      
\end{figure}

Referring to Figure \ref{fig:edges} the length of edges of a Voronoi cell at low densities can be computed solving the equations
\begin{eqnarray}
x^2 + \left(\frac{d}{2}\right)^2 &=& z^2 \\
y^2 + \left(\frac{b}{2}\right)^2 &=& z^2 \\
z &=& h-y\\
d^2 &=& h^2 + \left(\frac{b}{2}\right)^2\\
h &=& \frac{1}{\rho b}
\end{eqnarray}
for $x$, and $y$ depending on $b$ and $\rho$. Equally referring to Figure \ref{fig:edges} the length of edges of a Voronoi cell at low densities can be computed solving the equations
\begin{eqnarray}
w^2 &=& v^2+h^2\\
w^2 &=& u^2+\left(\frac{d}{2}\right)^2\\
w &=& \frac{b}{2}-v
\end{eqnarray}
for $u$, and $v$ depending on $b$ and $\rho$. The edge lengths and weights for ''Circum are shown in Figure \ref{fig:edgesweights}.

For ''Circum"' weighing the fundamental diagrams result for low and high density range as

\begin{eqnarray}
j(k<=1) &=& \frac{2k}{b^2} v_0 \left(1 -  \frac{\alpha}{2\sqrt{1+k^2}+\frac{1}{k} - k} \left( \frac{4}{\sqrt{1+k^2}} e^{-\frac{1}{2}\frac{b}{B} \sqrt{1+\frac{1}{k^2}}} \right)\right) \label{eq:flowlow}\\
j(k>1) &=& \frac{2k}{b^2} v_0 \left(1 -  \frac{\alpha}{2\sqrt{1+k^2}+k^2-1}\left( \frac{4}{\sqrt{1+k^2}} e^{-\frac{1}{2}\frac{b}{B} \sqrt{1+\frac{1}{k^2}}} + (k^2-1)e^{-\frac{b}{B}\frac{1}{k}}\right)\right)  \label{eq:flowhigh}
\end{eqnarray}

\begin{figure}[ht!]
\centering
\includegraphics[width=0.45\textwidth]{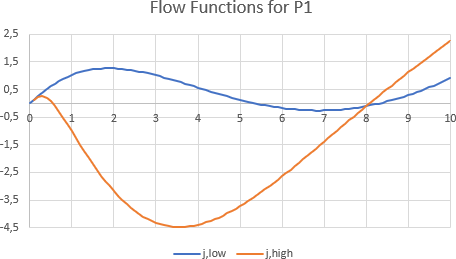} \hspace{11pt}
\includegraphics[width=0.45\textwidth]{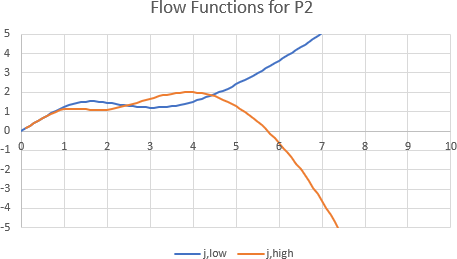}\\ \vspace{11pt}
\includegraphics[width=0.45\textwidth]{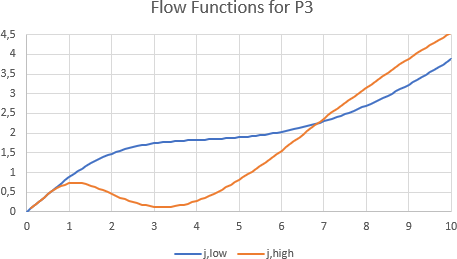} \hspace{11pt}
\includegraphics[width=0.45\textwidth]{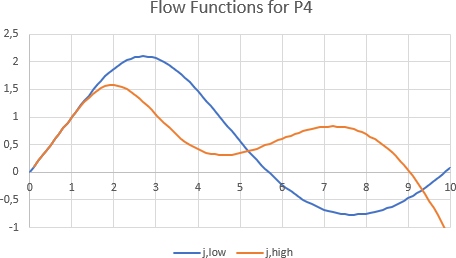}\\ \vspace{11pt}
\includegraphics[width=0.45\textwidth]{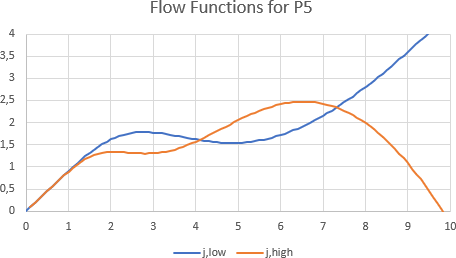} \hspace{11pt}
\caption{Functions for flow at low and high density, plotted over the full density range with the parameter sets which approximate the fundamental diagrams by Weidmann (P1), Hankin-Wright (P2), L{\"o}hner et al. (P3), Jin et al. (P4), and Predtechenskii and Milinskii (P5) from Table \ref{tab:parameters}.} 
\label{fig:flowlowhigh}      
\end{figure}

For illustration Figure \ref{fig:flowlowhigh} shows both functions (\ref{eq:flowlow}) and (\ref{eq:flowhigh}) with the parameters from Table \ref{tab:parameters}.

\begin{figure}[ht!]
\centering
\includegraphics[width=0.45\textwidth]{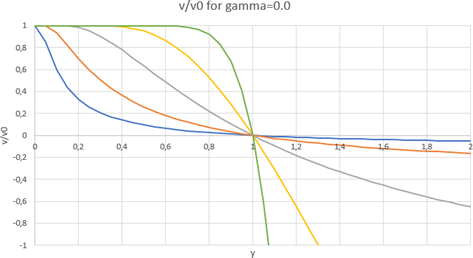} \hspace{11pt}
\includegraphics[width=0.45\textwidth]{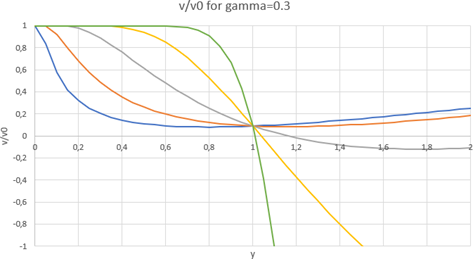}\\ \vspace{11pt}
\includegraphics[width=0.45\textwidth]{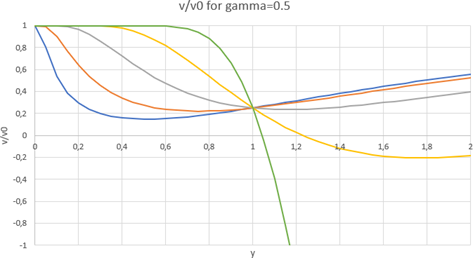} \hspace{11pt}
\includegraphics[width=0.45\textwidth]{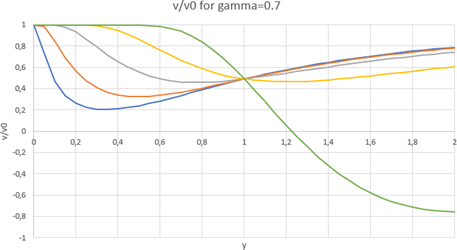}\\ \vspace{11pt}
\includegraphics[width=0.45\textwidth]{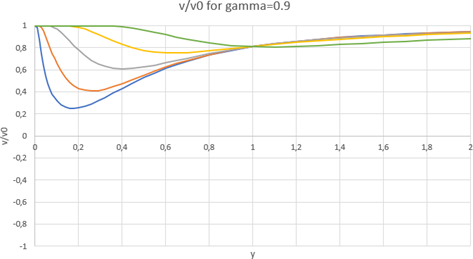} \hspace{11pt}
\includegraphics[width=0.45\textwidth]{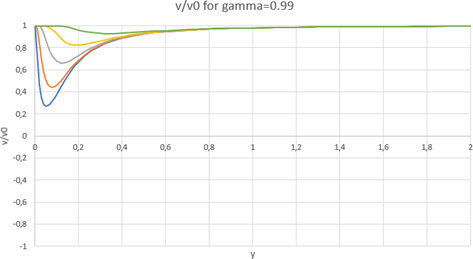}\\ \vspace{11pt}
\includegraphics[width=0.62\textwidth]{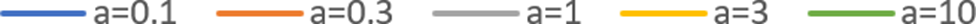}
\caption{Functions $f_{\gamma, a}(y)$ with values for $\gamma$ 0.0, 0.3, 0.5, 0.7, 0.9, and 0.99 and five different values for $a$.} 
\label{fig:fgammaa}      
\end{figure}

\begin{figure}[ht!]
\centering
\includegraphics[width=0.45\textwidth]{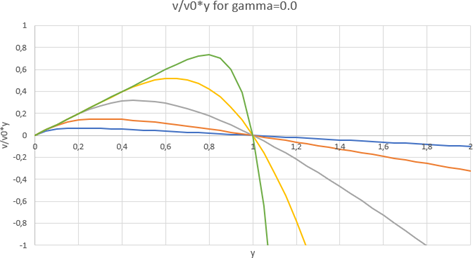} \hspace{11pt}
\includegraphics[width=0.45\textwidth]{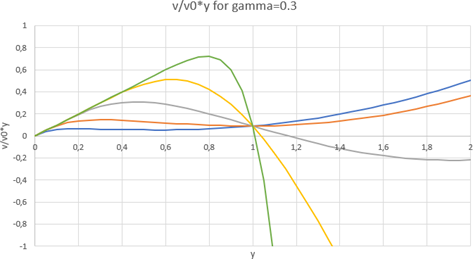}\\ \vspace{11pt}
\includegraphics[width=0.45\textwidth]{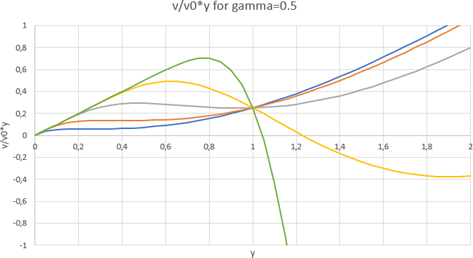} \hspace{11pt}
\includegraphics[width=0.45\textwidth]{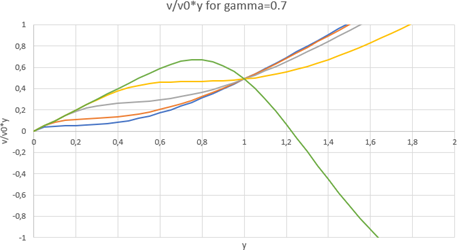}\\ \vspace{11pt}
\includegraphics[width=0.45\textwidth]{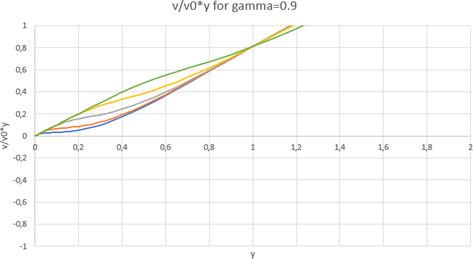} \hspace{11pt}
\includegraphics[width=0.45\textwidth]{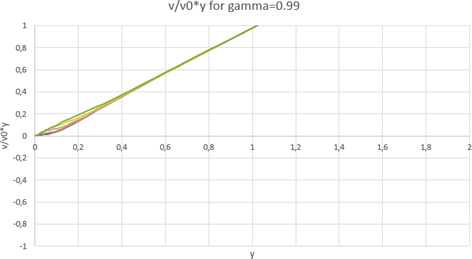}\\ \vspace{11pt}
\includegraphics[width=0.62\textwidth]{supplemental/a.png}
\caption{Functions $yf_{\gamma, a}(y)$ with values for $\gamma$ 0.0, 0.3, 0.5, 0.7, 0.9, and 0.99 and five different values for $a$.} 
\label{fig:yfgammaa}      
\end{figure}

It is possible to have a different perspective on the function for lower densities -- equation (\ref{eq:fd}) -- with a reparametrization:
\begin{eqnarray}
f_{\gamma, a}(y) &=& \frac{v}{v_0} = 1 - \frac{1-\gamma^2}{1-\gamma^2+\gamma^2y^2} e^{-a\left(\sqrt{\gamma^2+\frac{1-\gamma^2}{y^2}}-1\right)}\\
a&=&\ln\left(2(1-\lambda)A\frac{\tau}{v_0}\right)\\
\gamma &=& \frac{b}{2Ba}\\
y&=&\sqrt{\frac{1}{\gamma^2}-1} k
\end{eqnarray}
For $\gamma \rightarrow 0$ $f(y)$ approaches the function of the fundamental diagram which was used by Newell, Kladek, and Weidmann with a maximum density $y_{max}=1$ at which $f(y_{max})=0$. For larger values of $\gamma$ functions result at which $f(y=1)>0$. There are even functions $f_{\gamma, a}$ without a zero as figure \ref{fig:fgammaa} shows.
Looking at the corresponding flow functions $yf(y)$ in Figure \ref{fig:yfgammaa} one can see that not only the functional form of the Newell fundamental diagram is included ($\gamma \rightarrow 0$, $a \approx 0.3$), but also the one by L{\"o}hner et al. ($\gamma \approx 0.7$, $a \approx 1$). However, other functional forms from Figure \ref{fig:PFD01} have no counterpart.
These missing counterparts are only added with the change in the set of Voronoi neighbors when density increases.

\section{Some additional Discussion, Limitations, and Conclusions}
The extensive simplifications and assumptions (homogeneous population, infinitely large system) put a question mark at the relevance of the result. It could be a coincidence in the sense that analytical treatment does not cover the essential elements of a real system of pedestrians. In this case the obtained function would nevertheless be a candidate to be fitted to empirical results.

Some odd issues from the empirical data could not be clarified in this work. One of these is, why the functional form of fundamental diagrams of school boys and commuters is so similar at Hankin and Wright, whereas different functional forms are reported for commuters at stations by other researchers. Or why does the fundamental diagram by Helbing et al. fall into a different class than the one by L{\"o}hner et al. which was also obtained in Maccah, whereas the other fundamental diagrams in these two classes were reported from situations very different from the hajj. Such issues still make it seem possible that the difference between empirical pedestrian fundamental diagrams is caused entirely by different measurement techniques, data aggregation and fitting methods. If this should be shown to be the case one the, the starting point of this contribution would be invalidated.

For those fundamental diagrams which have their capacity at relatively high densities it was not possible to find parameters which do not overestimate capacity. This is probably not a serious problem since this could likely be explained with the assumption of a perfectly homogeneous population and that variations in the population produce a kind of friction which reduces flows, particularly maximum flow.

As much as the assumption of an infinitely large system is unrealistic the result of this work puts a question mark behind the relevance of single-file movement experiments of which have been popular over recent years, since the result put forth in this work emphasizes the relevance of dimensionality and that fundamental diagrams from single-file movement cannot be expanded to two dimensions with a simple global scaling factor.

An extension of the proposed model would be to consider more neighbors than only the nearest, with a neighborhood degree dependent suppression factor. An open question for this would be how to define the weight factors for non-nearest neighbors.

A hope which regrettably did not realize is to obtain a function to compute model parameters directly from basic empirical results. For this the function for the fundamental diagram is simply too complex. However, a -- for the purpose of calibration -- beneficial property that was observed previously is maintained: the possibility for step-wise calibration. First one would calibrate parameters $B$ and $\alpha$ with data from uni-directional movement and then calibrate the factors of $\alpha$ with data from more complex walking situations.

\end{document}